\documentclass{article}
\usepackage{spconf,amsmath,epsfig}
\usepackage{multirow}
\usepackage{dsfont}
\usepackage{stfloats}
\usepackage{subfig} 
\usepackage{tabu}  
\usepackage{cite}
\usepackage{booktabs}
\usepackage{algorithm}
\usepackage{algorithmic}
\usepackage{hyperref}

\title{Change Detection Between Optical Remote Sensing Imagery and Map Data via Segment Anything Model (SAM)}
\name
 {Hongruixuan Chen$^{1,2}$, 
 Jian Song$^{1,2}$, 
 Naoto Yokoya$^{1,2}$\sthanks{Corresponding Author (yokoya@k.u-tokyo.ac.jp). This work was supported in part by the Council for Science, Technology and Innovation (CSTI), the Cross-ministerial Strategic Innovation Promotion Program (SIP), Development of a Resilient Smart Network System against Natural Disasters (Funding agency: NIED), the JSPS, KAKENHI under Grant Number 22H03609, JST, FOREST under Grant Number JPMJFR206S, Microsoft Research Asia, Next Generation Artificial Intelligence Research Center of The University of Tokyo, and the Graduate School of Frontier Sciences, The University of Tokyo through the Challenging New Area Doctoral Research Grant (Project No. C2303).}
}
 
 \address{$^1$ Graduate School of Frontier Sciences, The University of Tokyo, Japan \\
 $^2$ RIKEN Center for Advanced Intelligence Project (AIP), RIKEN, Japan}
 
\begin{document}
%
\maketitle
\begin{abstract}
Unsupervised multimodal change detection is pivotal for time-sensitive tasks and comprehensive multi-temporal Earth monitoring. In this study, we explore unsupervised multimodal change detection between two key remote sensing data sources: optical high-resolution imagery and OpenStreetMap (OSM) data. Specifically, we propose to utilize the vision foundation model Segmentation Anything Model (SAM), for addressing our task. Leveraging SAM's exceptional zero-shot transfer capability, high-quality segmentation maps of optical images can be obtained. Thus, we can directly compare these two heterogeneous data forms in the so-called segmentation domain. We then introduce two strategies for guiding SAM's segmentation process: the ‘no-prompt’ and ‘box/mask prompt’ methods. The two strategies are designed to detect land-cover changes in general scenarios and to identify new land-cover objects within existing backgrounds, respectively. Experimental results on three datasets indicate that the proposed approach can achieve more competitive results compared to representative unsupervised multimodal change detection methods. 
\end{abstract}
\section{Introduction}
\label{sec:intro}

\par Multimodal change detection aims at detecting land-cover changes from multitemporal remote sensing images with different modalities \cite{Touati2020a, Chen2019a}. As leveraging various types of data sources, this technique is crucial for high temporal resolution monitoring and rapid response to emergent events \cite{Sun2021a, Chen2023Fourier, Yi2024TTST, Hu2023Binary}. However, compared to unimodal change detection, multimodal change detection presents additional challenges due to variations in statistical distributions, channel numbers, and noise levels between pre-event and post-event images, known as the heterogeneous modality problem \cite{Chen2023Fourier}. 

\par Depending on whether or not labels are available to change detectors, existing approaches are categorized into supervised, semi-supervised, and unsupervised methods. Among them, unsupervised methods have gained prominence due to the fact that detectors can be used without any prior-labeled data. The basic idea of unsupervised multimodal change detection is to transform multimodal images into a new domain where the modal heterogeneity can be eliminated. Depending on the transformation technique and the target domain, the existing unsupervised methods can be categorized into four types, i.e., (1) modality translation-based methods \cite{Mignotte2020}, (2) similarity measurement-based methods \cite{Sun2021a, Sun2021c, chen2022unsupervised, Chen2023Fourier}, (3) feature learning-based methods \cite{Liu2018}, and (4) classification-based methods \cite{Wan2019Post}. 

\par Currently, the multimodal data involved in most of these methods are remote sensing images acquired by airborne and spaceborne sensors. Few studies have focused on map data, such as OpenStreetMap (OSM) data. Multimodal change detection between map data and optical imagery is undoubtedly significant in enriching the data sources of change detection as well as updating the geographic information system. Our previous work \cite{chen2023land} explored this by developing an architecture to detect land-cover changes between OSM data and optical high-resolution imagery and established the first benchmark dataset based on the OpenEarthMap dataset \cite{Xia2023OpenEarthMap}. However, the proposed architecture and focused tasks are still supervised. How to achieve change detection between map data and optical imagery in an unsupervised manner remains an unexplored and challenging topic. Existing unsupervised models, designed for airborne and spaceborne remote sensing images, may struggle with the unique aspects of map data. Therefore, it may be difficult for them to achieve good detection results meeting real-world application requirements. In this paper, we advance the field by proposing an unsupervised approach to detect land-cover changes between map data and optical imagery through exploiting the vision foundation model SAM \cite{Kirillov2023Segment}. 
 
\section{Methodology}\label{sec:2}
\subsection{Segment Anything Model}

\par As a vision foundation model, the SAM is designed and trained to be promptable \cite{Kirillov2023Segment}. By trained on the millions of annotated images, the SAM can perform zero-shot transfer for new image distributions and tasks. The architecture of SAM comprises three key components: an image encoder, a prompt encoder that can receive points, boxes, text, and mask as the prompt, and a lightweight mask decoder. Among them, the image encoder is a pre-trained Vision Transformer (ViT) \cite{dosovitskiy2020image}; the positional encodings, CLIP model \cite{Radford2021Learning} and convolutional layers are adopted to embed different types of prompt data; the decoder integrates a Transformer decoder block with a dynamic mask prediction head. Now, SAM has demonstrated impressive performance in a range of remote sensing applications \cite{OSCO2023Segment}. 

\par In this paper, we achieve unsupervised change detection on optical and map data by leveraging the powerful zero-shot segmentation capability of SAM to transform the optical image into the modality-independent segmentation domain.

\subsection{Detecting Land-Cover Changes}
\begin{figure}[t]
  \hfill
  \subfloat[]{
    \includegraphics[width=3.32in]{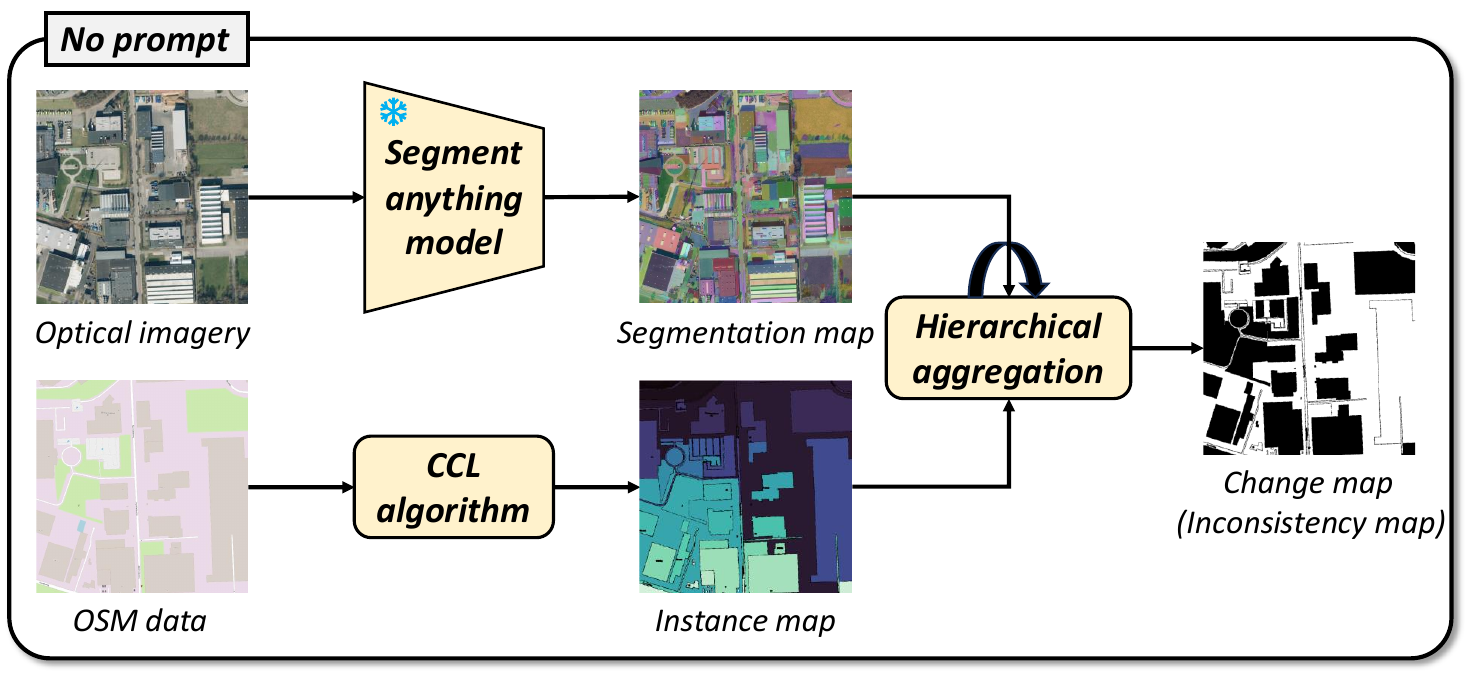}
  \label{fig_first_case}}
  \hfil
  \subfloat[]{
    \includegraphics[width=3.32in]{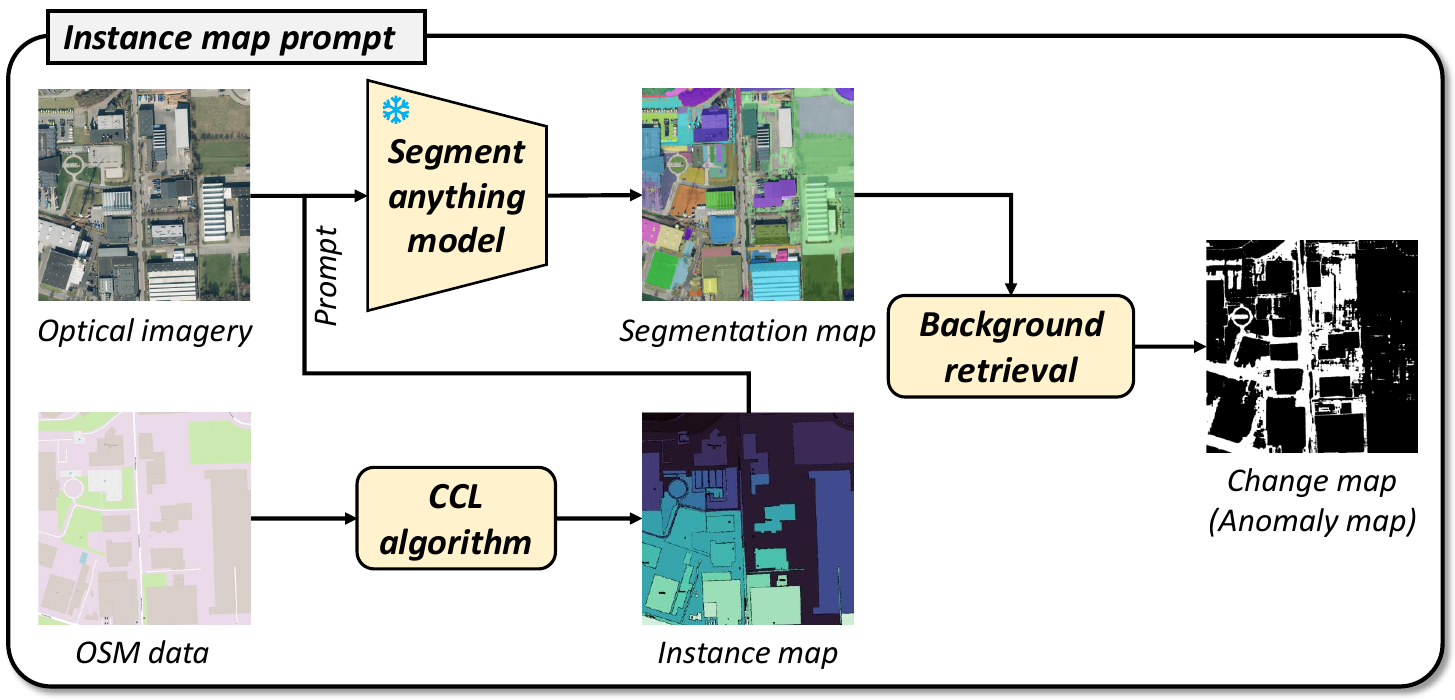}
  \label{fig_second_case}}
  \caption{The proposed multimodal change detection framework based on the SAM. (a) Detecting land-cover changes without prompt by comparing the shape of instances. (b) Detecting land-cover changes with the prompt from instance maps. }
  \label{Fig:SAMMCD_framework}
\end{figure}

\par We propose two strategies to detect land-cover changes using the SAM. The first strategy detects land-cover changes in the general case, while the second targets the detection of new land-cover objects appearing against a background.

\subsubsection{No Prompt}

\par As illustrated in Fig. \ref{Fig:SAMMCD_framework}-(a), the first approach employs SAM's no-prompt function, i.e., let SAM segment everything in the optical imagery to generate its segmentation map. Concurrently, we apply a connected component labeling (CCL) algorithm on the rasterized OSM data to produce its instance map. This process aligns the optical image and OSM data within the same domain, which we call the segmentation domain, thus eliminating the modality differences between them. Subsequently, in order to detect land-cover changes from the obtained segmentation and instance maps, here, we argue that two land-cover objects should have different shapes if a change event occurs. In this way, land-cover changes can be obtained by comparing the shape attributes such as area, aspect ratio, etc. of two instances at the same location. 

\par However, the masks obtained by the SAM do not have the category information and each mask may not represent a complete land-cover instance. For example, a building might be composed of several masks. Therefore, we propose a hierarchical aggregation method guided by OSM data instances. Specifically, for each instance in the OSM data, we find all masks in the segmentation map with which there is an intersection, and iteratively merge these masks outward from the instance's center. After each merge operation, we calculate the overlap rate between the merged mask and the instance. If the overlap rate exceeds a set threshold during the merge process, the instance is considered to be unchanged, or else it is considered to be changed. 

\subsubsection{Instance Map Prompt}

\par The above strategy can effectively detect the situation in which two instances have changed. However, it struggles with scenarios where new land-cover objects emerge within a land-cover background. For example, a certain large area is vegetation on the OSM data, and a building appears in the optical image. In this case, the above strategy treats the entire background area as unchanged and the emerging building cannot be detected.

\par To address this limitation, we propose a strategy that adopts instances from OSM data as prompts for SAM. As depicted in Fig. \ref{Fig:SAMMCD_framework}-(b), we can guide the segmentation of SAM from using background instances of OSM data, which can be obtained from the legend of the OSM data \cite{chen2023land}, in the form of a box or mask prompt. In this case, SAM will generally segment the background in the optical image. The land-cover objects appearing in the background will be considered anomalies, and thus will not be segmented out by SAM. In this way, we can identify these emerging objects by extracting unrecognized pixels from the segmentation map within the instance's region.

\section{Experiments}
\label{sec:experiments}
\subsection{General Information}\label{sec:general}
\par In this paper, our experiments utilize three datasets: Aachen, Christchurch, and Vegas \cite{chen2023land}. These datasets vary in size: Aachen measures 1000$\times$1000 pixels, Christchurch 1024$\times$1024 pixels, and Vegas 650$\times$650 pixels. As illustrated in Fig. \ref{fig:Aachen_bcm} through Fig. \ref{fig:Vegas_bcm}, the substantial modality differences between map data and optical remote sensing imagery present significant challenges for change detection.

\par We have selected six prominent unsupervised multimodal change detection methods for comparative analysis. These methods are M3CD \cite{Touati2020a}, FPMS \cite{Mignotte2020}, NPSG \cite{Sun2021c}, IRG-McS \cite{Sun2021a}, SR-GCAE \cite{chen2022unsupervised}, and FD-MCD \cite{Chen2023Fourier}. Each has demonstrated state-of-the-art results across various benchmark datasets and modal combinations.

\par To assess the accuracy of these methods, we employ three commonly used metrics in change detection tasks \cite{Chen2023Exchange}: overall accuracy (OA), F1 score, and Kappa coefficient (KC). These metrics will provide a comprehensive evaluation of each method's performance.

\subsection{Experimental Results}\label{sec:experi_result}
\begin{figure}[!t]
  \centering
\includegraphics[width=3.3in]{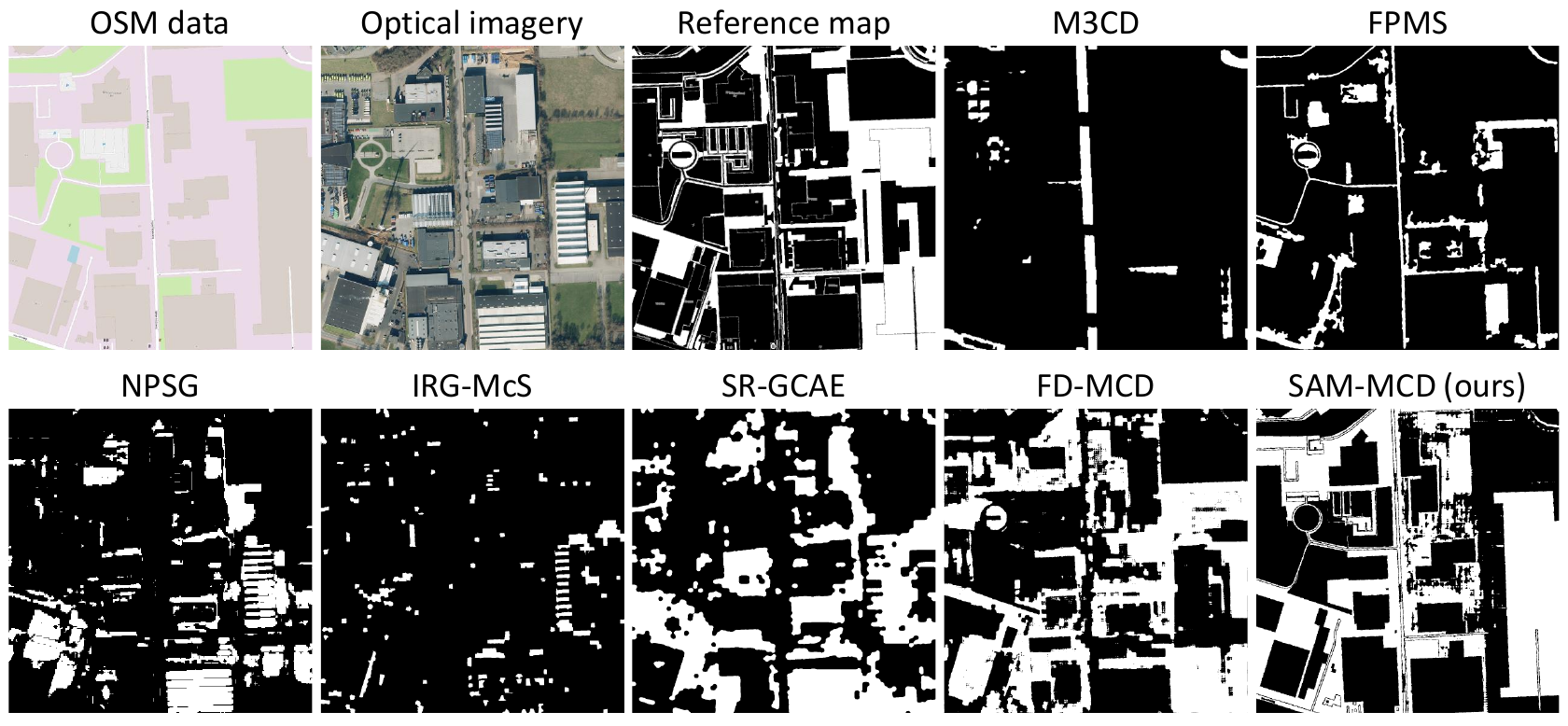}
  \caption{Binary change maps obtained by different methods on the Aachen dataset. }
  \label{fig:Aachen_bcm}
\end{figure}

\begin{figure}[!t]
  \centering
\includegraphics[width=3.3in]{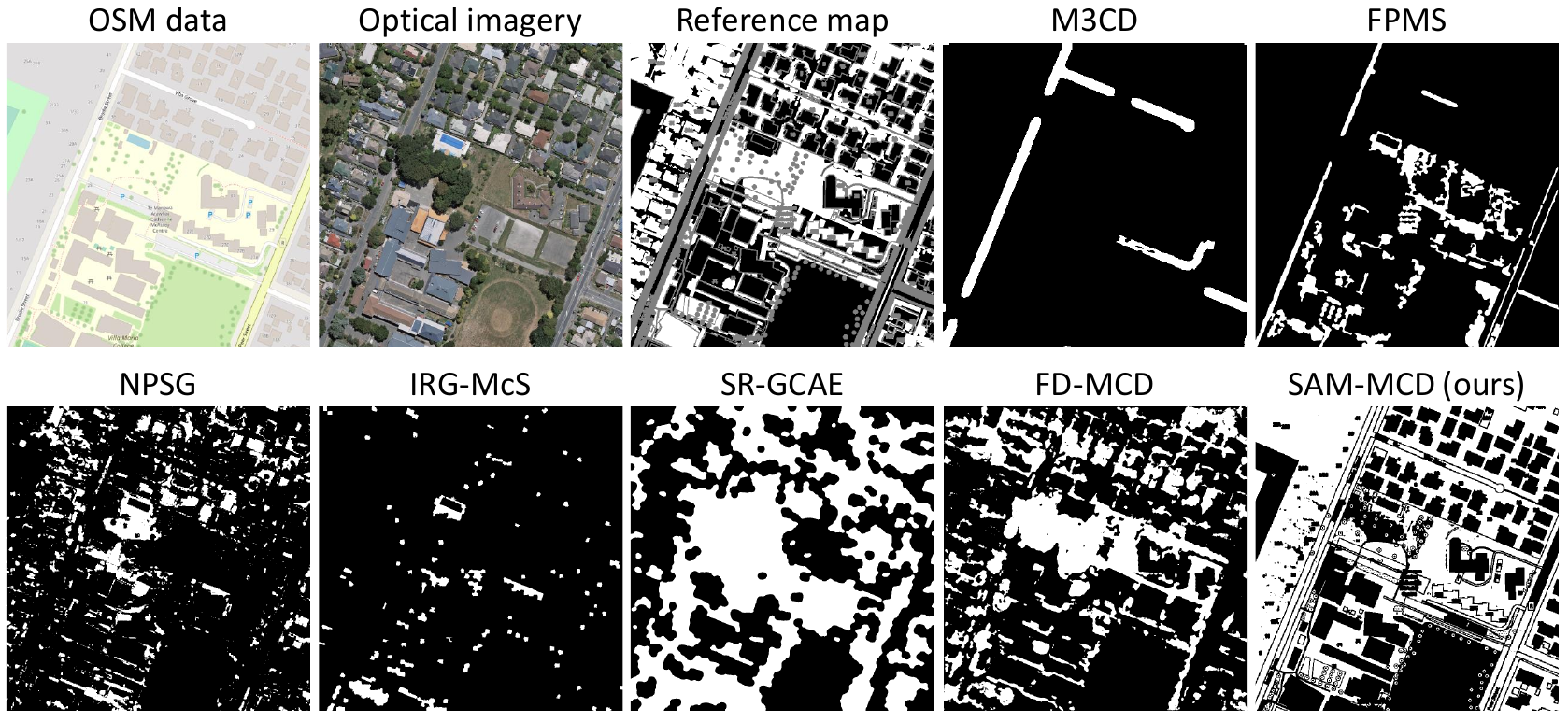}
  \caption{Binary change maps obtained by different methods on the Christchurch dataset. }
  \label{fig:Christchurch_bcm}
\end{figure}

\begin{figure}[!t]
  \centering
\includegraphics[width=3.3in]{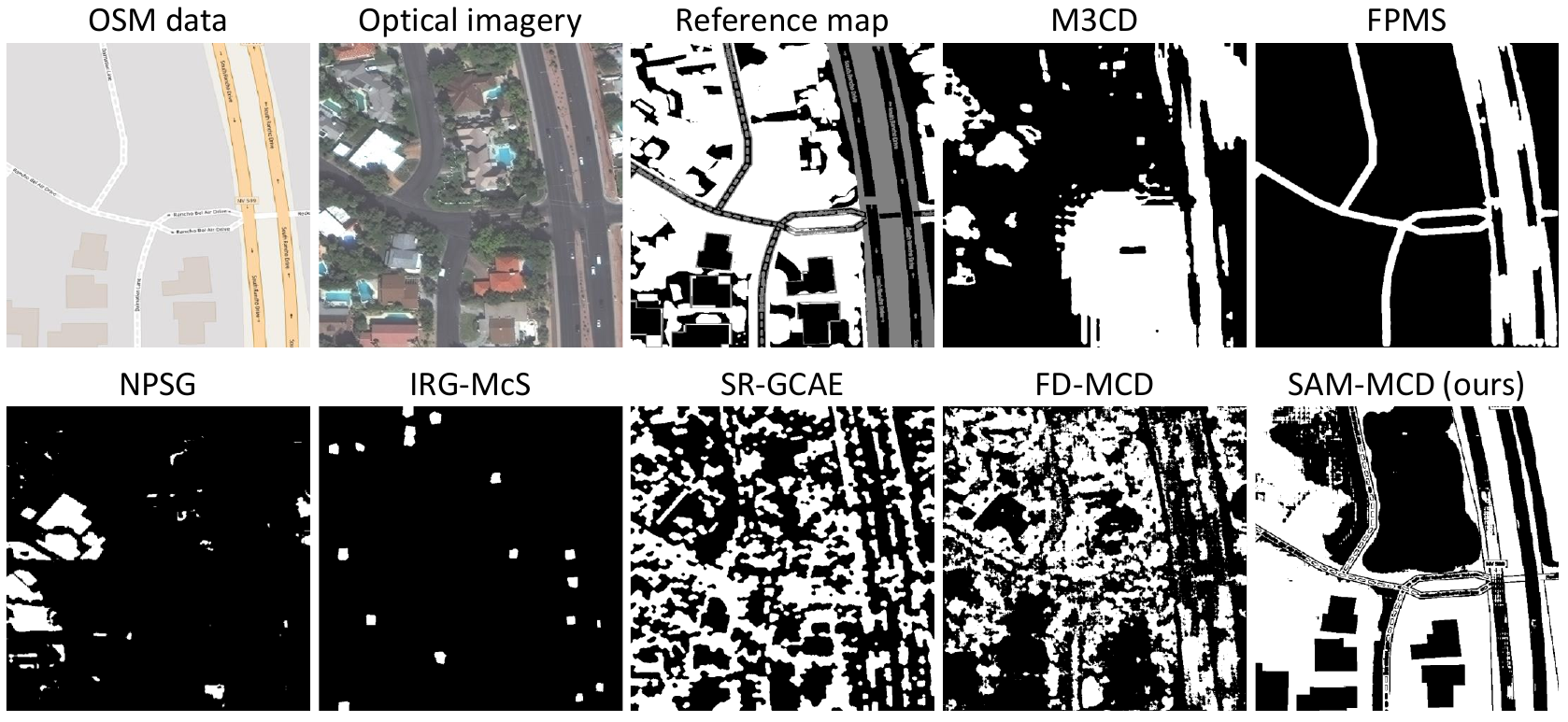}
  \caption{Binary change maps obtained by different methods on the Vegas dataset. }
  \label{fig:Vegas_bcm}
\end{figure}

\begin{table*}[!ht]
  \setlength\tabcolsep{5pt} \small
  \renewcommand{\arraystretch}{1.2}
  \caption{\centering{Accuracy assessment on change maps obtained by different methods on the three multimodal change detection datasets. The highest values are marked in \textbf{bold} and the next highest values are \underline{underlined}.}}
  \label{tbl:acc_ass}
  \centering
  \begin{tabular}{c |c c c |c c c | c c c}
    \hline			
    \multirow{2}{*}{\textbf{Method}} & \multicolumn{3}{c}{\textbf{Aachen}} & \multicolumn{3}{c}{\textbf{Christchurch}} & \multicolumn{3}{c}{\textbf{Vegas}} \\
    \cline{2-10} 
    & OA & F1 & KC & OA & F1 & KC & OA & F1 & KC  \\
    \hline\hline
    M3CD \cite{Touati2020a} & 0.6542 & 0.0946 & 0.0083 & 0.3790  & 0.0582  & -0.0708 & 0.4485 & 0.4233 & -0.0141 \\ 	
    FPMS \cite{Mignotte2020} & \textbf{0.6954} & 0.3233 & 0.1804 &  0.5269 & 0.1800 & 0.0100 & 0.3184 & 0.0076 & -0.1126 \\ 
    NPSG \cite{Sun2021c} &  0.6376 & 0.3051 & 0.0840 &  0.5149 & 0.2186 &  -0.0099 &  0.4225&  0.1805 &  0.0503\\ 
    IRG-McS \cite{Sun2021a} &  0.6780  &  0.1617 & 0.0786 & 0.5272 & 0.0638 &  0.0020 & 0.3788 &  0.0364 &  0.0035 \\ 
    SR-GCAE \cite{chen2022unsupervised}  &  0.6136 & 0.4344  & 0.1414    &  0.5656 & 0.5228 & 0.1252 & 0.5026 & 0.5179& 0.0517 \\ 
    FD-MCD \cite{Chen2023Fourier} &  0.6684 & \underline{0.5540} & \underline{0.2949}  &   \underline{0.7449} &  \underline{0.6737} &  \underline{0.4787}  &  \textbf{0.5792} & \underline{0.6017} & \textbf{0.1878}  \\ 
    SAM-MCD &  \underline{0.6820} & \textbf{0.5945} &  \textbf{0.3429}  & \textbf{0.7503}  & \textbf{0.7729} &  \textbf{0.5077} & \underline{0.5735}  & \textbf{0.6251} &  \underline{0.1428} \\ 

    \hline
  \end{tabular}
\end{table*}

\par Figures \ref{fig:Aachen_bcm} through \ref{fig:Vegas_bcm} display binary change maps generated by the comparison methods and our proposed SAM-MCD. Specific accuracy metrics for these methods are detailed in Table \ref{tbl:acc_ass}. For most comparison methods, although they can obtain accurate detection results on data from the modal combinations like optical and SAR data pairs, their effectiveness diminishes when detecting land-cover changes between map and optical data pairs. Among all comparison methods, FD-MCD \cite{Chen2023Fourier}, which analyzes structural relationships by transforming data from different modalities into the (graph) Fourier domain, shows commendable performance across all three datasets.

\par In comparison, the proposed method obtains more accurate detection results by leveraging the advanced segmentation capabilities of the vision foundation model and comparing the relationship between the map data and the optical images in the modality-independent segmentation domain. However, it is observed that our method encounters some challenges with the Vegas dataset, particularly in missing detections. This issue predominantly arises in certain regions of the optical image (notably the upper center region), where our two strategies are less effective. Consequently, our method fails to detect most of the changes occurring in these areas.

\section{Conclusion}
\label{sec:conclusion}

\par In this paper, we propose an unsupervised multimodal change detection approach aiming at automatically detecting land-cover changes from OSM data and optical remote sensing imagery. By employing the vision foundation model SAM, OSM data and optical images with large modality difference can be transformed to the modality-independent segmentation domain. We design two strategies for detecting changes based on segmentation maps of optical images and instance maps of OSM. When applied to three map-optical data pairs with distinct scenes, the proposed method yields more competitive detection results compared to the SOTA methods. Our future work will focus on extending our framework to unsupervised semantic change detection on OSM data and optical imagery.


\small
\bibliographystyle{IEEEbib}
\bibliography{strings,refs}

\end{document}